\documentclass[aps,pra,twocolumn,10pt,showpacs,superscriptaddress,floatfix,nofootinbib]{revtex4-2}
\usepackage{float}
\usepackage{graphicx}
\usepackage{amsmath}
\usepackage{amssymb}
\usepackage{tabularx}
\usepackage{multirow}
\usepackage{xifthen}
\usepackage{tikz}
\usepackage[dvipsnames]{xcolor}
\usepackage[export]{adjustbox}

\newcommand{\Matrix}[1]{\begin{Vmatrix}#1\end{Vmatrix}}
\newcommand{\Gate}[4]{\textsf{#1}_{#3}^{#4}\ifthenelse{\isempty{#2}}{}{\left(#2\right)}}
\newcommand{\Utry}[1][]{\Gate{U}{}{#1}{}}
\newcommand{\UtryElt}[2]{\mathcal{U}_{#1,#2}}
\newcommand{\Id}{\Gate{Id}{}{}{}}
\newcommand{\Ph}[2][]{\Gate{P}{#1}{}{#2}}
\newcommand{\Rp}[2][]{\Gate{R}{#1}{}{#2}}
\newcommand{\Sw}[1]{\Gate{S}{}{}{#1}}
\newcommand{\BS}[2][]{\Gate{T}{#1}{}{#2}}

\tikzstyle{level-node}=[
    circle,
    draw=black,
    fill=blue!10,
    thick,
    minimum size=1cm,
]

\tikzstyle{trans-edge}=[
    {Latex[length=.2cm]}-,
]

\usepackage[bookmarks=true,colorlinks=true,linkcolor=blue,urlcolor=blue,citecolor=blue,bookmarks=true,hyperindex=true]{hyperref}
\usepackage{caption}
\usepackage{subcaption}
\usepackage{cleveref}
\usepackage{listings}
\usepackage{braket}
\usepackage{orcidlink}

\Crefformat{lstlisting}{Listing~(#2#1#3)}

\newcommand{\TO}{\textbf{TO}}

\begin{document}

\title{Transition-Aware Decomposition of Single-Qudit Gates}

\author{Denis A. Drozhzhin\,\orcidlink{0009-0006-3114-4543}}
\author{Evgeniy O. Kiktenko\,\orcidlink{0000-0001-5760-441X}}
\author{Aleksey K. Fedorov\,\orcidlink{0000-0002-4722-3418}}
\author{Anastasiia S. Nikolaeva\,\orcidlink{0000-0002-8321-7103}}
\affiliation{Laboratory of Quantum Information Technologies, National University of Science and Technology ``MISIS'',  Moscow 119049, Russia}

\begin{abstract}
Quantum computation with $d$-level quantum systems, also known as qudits, benefits from the possibility to use a richer computational space compared to qubits.
However, for an arbitrary qudit-based hardware platform, the issue is that a generic qudit operation has to be decomposed into the sequence of native operations---pulses that are adjusted to the transitions between two levels in a qudit.
Typically, not all levels in a qudit are simply connected to each other due to specific selection rules.
Moreover, the number of pulses plays a significant role, since each pulse takes a certain execution time and may introduce error.
In this paper, we propose a resource-efficient algorithm to decompose single-qudit operations into the sequence of pulses that are allowed by qudit selection rules.
Using the developed algorithm, the number of pulses is at most $d(d{-}1)/2$ for an arbitrary single-qudit operation.
For specific operations, the algorithm could produce even fewer pulses.
We provide a comparison of qudit decompositions for several types of trapped ions, specifically $^{171}\text{Yb}^+$, $^{137}\text{Ba}^+$ and $^{40}\text{Ca}^+$ with different selection rules, and also decomposition for superconducting qudits.
Although our approach deals with single-qudit operations, the proposed approach is important for realizing two-qudit operations since they can be implemented as a standard two-qubit gate that is surrounded by efficiently implemented single-qudit gates.
\end{abstract}


\maketitle

\section{Introduction}

The development of a large-scale quantum computer~\cite{Brassard1998,Ladd2010,Fedorov2022} that is able to outperform devices based on classical principles in practically relevant problems, such as simulating complex (quantum) systems~\cite{Lloyd1996}, combinatorial optimization~\cite{Farhi2014}, or prime factorization~\cite{Shor1994}, represents an outstanding challenge.
Various computational models, for example, digital quantum computing~\cite{Brassard1998,Ladd2010}, variational quantum algorithms~\cite{Babbush2021-4,Aspuru-Guzik2022}, or programmable quantum simulation~\cite{Lukin2021}, and diverse physical platforms,
such as superconducting circuits~\cite{Martinis2019,Pan2021-4}, semiconductor quantum dots~\cite{Vandersypen2022,Morello2022,Tarucha2022}, optical systems~\cite{Pan2020,Lavoie2022}, neutral atoms~\cite{Lukin2021,Browaeys2021,Browaeys2020-2,Saffman2022}, and trapped ions~\cite{Monroe2017,Blatt2012,Blatt2018}, are under study on the way to useful quantum computers.
The inherent idea is to find suitable conditions for scaling quantum devices with respect to the number of available information carriers without significantly degrading the quality of the control.
In the most well-developed digital model of quantum computing, information carriers are qubits~\cite{Brassard1998,Ladd2010,Fedorov2022}, which are two-level quantum counterparts of classical bits.
The execution of quantum algorithms requires the realization of single- and two-qubit operations under a register of qubits, so that the combination of the number of qubits and quality of quantum operations (gates) is the crucial parameter.

Underlying physical platforms, such as trapped ions or neutral atoms, however, almost always have Hilbert spaces of higher dimensionalities, which gives interesting possibilities~\cite{Kiktenko2023rmp}.
Computing models based on the use of additional degrees of freedom of physical systems, which makes them qudits (where $d$ indicates that the dimension of the Hilbert space may be larger than two), attract great interest since they allow scalability of quantum computing devices without the need for an increase in the number of physical carriers~\cite{Farhi1998,Kessel1999,Kessel2000,Kessel2002,Muthukrishnan2000,Nielsen2002,Berry2002,Klimov2003,Bagan2003,Vlasov2003,Clark2004,Leary2006,Ralph2007,White2008,Ionicioiu2009,Martinis2009,White2009,Ivanov2012,Mischuck2012,Wallraff2012,Li2013,Martinis2014,Gustavsson2015,Kiktenko2015,Kiktenko2015-2,Ustinov2015,Song2016,Frydryszak2017,Balestro2017,Bocharov2017,Morandotti2017,Gokhale2019,Pan2019,Low2020,Jin2021,Sawant2020,Pavlidis2021,Rambow2021}.
There are several approaches to how qudits can be used for more efficient quantum processors.
First, one can think about a multilevel system with $2^n$ levels as a set of qubits: for example, a ququart is equivalent to a two-qubit system~\cite{Nikolaeva2021epj,Nikolaeva2023universal,Drozhzhin2024}.
At the same time, implementing a two-qubit operation with a single ququart (strictly speaking, this is a single-qudit operation) would not require physical interaction between physical objects,
so the fidelity of such operations can be as high as the fidelity of single-qubit operations.
The second approach is the use of additional qudit levels to substitute ancilla qubits in multiqubit gate decompositions~\cite{Barenco1995}
(for example, for the Toffoli gate~\cite{Ralph2007,White2009,Ionicioiu2009,Wallraff2012,Kwek2020,Baker2020,Kiktenko2020,Kwek2021,Galda2021,Gu2021}).
However, the efficiency of these methods and their combination depends on the mapping, i.e., the way in which qubits are encoded in qudits~\cite{Nikolaeva2021epj,Drozhzhin2024}.

As in the case of conventional qubit-based processors, the choice of a specific physical platform is also important for qudit setups.
Recently, multiqudit quantum processors~\cite{Hill2021,Wang2022,Ringbauer2022,Semerikov2022}, including systems based on superconducting transmon qudits~\cite{Hill2021} and photons~\cite{Wang2022} as well as $^{40}$Ca$^{+}$~\cite{Ringbauer2022} and $^{171}$Yb$^{+}$~\cite{Semerikov2022} ion qudits, have been demonstrated.
It is clear that in the case of atomic qubits (ions or neutral atoms), qudit encoding is natural due to multilevel structures, and it is straightforward to address more than two levels using a single laser with acousto-optic modulators (AOMs)~\cite{Ringbauer2022,Semerikov2022}.
Superconducting~\cite{Blok2025_8levels,Blok2025_12levels} and photonic~\cite{Wang2022_simulating} systems are also promising for qudit-based quantum computing.
Recent progress in quantum computing with qudits also includes the demonstration of a significant improvement in the realization of multiqubit gates~\cite{Nikolaeva2022,Nikolaeva2025ions} and quantum algorithms with qudits~\cite{Martinis2009,Blok2025_8levels}.
However, various details of resource-efficient implementation of quantum algorithms with qudits are the subject of research.

In this work, we address a specific problem of the realization of single-qudit operations, taking into account selection rules of physical systems.
The objective is to minimize the number of pulses in order to realize a single-qudit operation, since such a minimization would result in shorter execution time and a lower number of errors.
We propose a resource-efficient algorithm to decompose qudit operations into the sequence of pulses that are allowed by qudit selection rules.
As we demonstrate, in the proposed algorithm, the number of pulses is at most $d(d{-}1)/2$ for an arbitrary single-qudit operation, or even fewer pulses for specific single-qudit operations.
We provide a comparison of qudit decompositions for several types of trapped ions, specifically $^{171}\text{Yb}^+$, $^{137}\text{Ba}^+$ and $^{40}\text{Ca}^+$ with different selection rules, and also decomposition for superconducting qudits.

\section{Qudit Computing}
\label{section:qudit-logic}

A common quantum device relies on operations with \textit{qubits}, i.e., the set of transformations of two-level systems in the superposition states:
\begin{equation}
    \ket{\psi}_{\textbf{qb}} = \psi_0\ket{0} + \psi_1\ket{1}.
\end{equation}
In the case of \textit{qudits}, the structure of the superposition state of the $d$-level quantum system is more complex:
\begin{equation}
    \ket{\psi}_{\textbf{qd}} = \psi_0\ket{0} + \cdots + \psi_{d{-}1} \ket{d{-}1} = \sum_{n=0}^{d-1} \psi_{n}\ket{n},
\end{equation}
where $d$ is the dimension of the quantum state.

The evolution of a qudit state can be described using unitary operators, similarly to the case of qubit operators.
For example, if we interpret qubit Pauli $\sigma_x$ as an operation that swaps two levels, then we obtain:
\begin{equation}
    \Gate{SWAP}{}{\textbf{qd}}{ij} = \ket{i}\bra{j} + \ket{j}\bra{i} + \sum_{n\ne i,j} \ket{n}\bra{n}.
\end{equation}
On the other hand, $\sigma_x$ can be generalized to modulo $d$ increment:
\begin{align}
    \label{eqn:qudit-x-gate}
    \Gate{X}{}{\textbf{qd}}{} &= \sum_{n} \ket{n{+}1 \bmod d}\bra{n}, \\
    \Gate{X}{}{\textbf{qd}}{k} &= \sum_{n} \ket{n{+}k \bmod d}\bra{n}.
\end{align}

Also, one can use phase gates $\Gate{Z}{}{\textbf{qd}}{}$ and the $d$-nary Hadamard transform $\Gate{H}{}{\textbf{qd}}{}$, which is equivalent to the Quantum Fourier Transform on a single qudit:
\begin{align}
    \Gate{Z}{}{\textbf{qd}}{} &= \sum_{n} \omega^{n} \ket{n}\bra{n}, \\
    \label{eqn:qudit-h-gate}
    \Gate{H}{}{\textbf{qd}}{} &= \sum_{n,m} \frac{\omega^{nm}}{\sqrt{d}} \ket{n}\bra{m},
\end{align}
where $\omega = e^{\iota2\pi / d}$ is the $d^\text{th}$ root of unity.

In general, any single-qudit evolution can be described using elements of the unitary group $\Utry[\textbf{qd}]{} \in U(d)$.
This group is isomorphic to the group of $d{\times}d$ matrices with complex entries that satisfy the unitary condition:
\begin{equation}
    \Utry[\textbf{qd}]{}
    ~\equiv~
    \Matrix{
        \UtryElt{0}{0} & \UtryElt{0}{1} & \cdots & \UtryElt{0}{d-1} \\
        \UtryElt{1}{0} & \UtryElt{1}{1} & \cdots & \UtryElt{1}{d-1} \\
        \vdots & \vdots & \ddots & \vdots \\
        \UtryElt{d-1}{0} & \UtryElt{d-1}{1} & \cdots & \UtryElt{d-1}{d-1} \\
    }
\end{equation}
\begin{equation}
    \Utry[\textbf{qd}]{}^\dagger \Utry[\textbf{qd}]{} = \Id
    ~\equiv~
    \sum_{k=0}^{d-1} \UtryElt{k}{n}^*~\UtryElt{k}{m} = \delta_{n,m}
\end{equation}
The knowledge of how to efficiently implement arbitrary single-qudit evolution with existing qudit hardware is the crucial step towards the universal qudit computer, especially for systems with high dimensionality \cite{Blok2025_8levels,Blok2025_12levels,Bullock2005,Low2025_13levels,Low2025_25levels}.

Another important part of computation with qudits is entangling operations.
Using the extension of Pauli operations to qudits, we obtain the analogues of qubit controlled gates, particularly $\Gate{CX}{}{}{}$ and $\Gate{CZ}{}{}{}$:
\begin{align}
    \Gate{CU}{}{\textbf{qd}}{} &= \sum_{m=0}^{d-1} \ket{m}\bra{m} \otimes \Gate{U}{}{\textbf{qd}}{m}, \\
    \Gate{CX}{}{\textbf{qd}}{} &= \sum_{m,n} \ket{m,n{+}m}\bra{m,n}, \\
    \Gate{CZ}{}{\textbf{qd}}{} &= \sum_{m,n} \omega^{mn} \ket{m,n}\bra{m,n}.
\end{align}
Also, on several trapped-ion platforms, the main two-qudit operation is M\o{}lmer-S\o{}rensen gate~\cite{Ringbauer2022,Zalivako2024qb16}:
\begin{equation}
    \Gate{MS}{\chi}{}{} = \exp\left\{-\iota\frac{\chi}{2} \left(\sigma^{01}_x \otimes 1 + 1 \otimes \sigma^{01}_x \right)^2\right\},
\end{equation}
which is equivalent up to global phase to qubit $\Gate{RXX}{}{}{}$ operation on the subspace $\left\{\ket{00},\ket{01},\ket{10},\ket{11}\right\}$:
\begin{equation}
    \Gate{RXX}{\chi}{}{} = \exp\left\{-\iota\chi~\sigma_x \otimes \sigma_x \right\}.
\end{equation}

Although we do not consider qudit entangling operations within the scope of this paper, we are able to embed $n$-qubit gates into a single-qudit operation if $d \ge 2^n$.

\section{Qudit Hardware Operations}
\label{section:qudit-hardware}

From an experimental point of view, the qudit system can be implemented using different physical systems: trapped ions, neutral atoms, superconducting circuits, photonic circuits, etc.

A qudit state can be encoded into ions' or neutral atoms' degrees of freedom, e.g., states of the electron on the outer shell~\cite{Low2025_13levels,Low2025_25levels,Zalivako2024qb16}.
In this model, each state $\ket{n}$ has the corresponding energy $E_n$.
After applying the laser field with a frequency $\frac{1}{\hbar}\left|E_i - E_j\right|$, the Rabi oscillation of the electron between states $\ket{i}$ and $\ket{j}$ is stimulated if this transition is permitted by selection rules.

Superconducting circuits are able to store qudit states as Fock states of Copper pairs in the potential well~\cite{Blok2025_8levels,Blok2025_12levels}.
Transitions between states are also stimulated via Rabi pulses between neighboring states $\ket{n}$ and $\ket{n{+}1}$.
The potential well should provide some degree of anharmonicity so that each transition can be addressed individually.

Photonic circuits can represent a qudit state using $d$-wire schemes with a single photon~\cite{Wang2022,Wang2022_simulating}.
An index of the basis qudit state represents the index of the wire.
Using the beam-splitter with a given amplitude reflection rate $r$ (it is usually set $r{=}1/\sqrt{2}$ for a half-transparent mirror), we can permute photons in two neighboring wires.
Also, we can perform a phase-shift of a single wire at an arbitrary angle.

We can note that every mentioned qudit platform supports two main types of operations.
The first is the phase shift on the level $\ket{k}$, which we denote $\Ph[\theta]{k}$.
The second is a two-level transition $\Rp[\theta,\phi]{ij}$ between levels $\ket{i}$ and $\ket{j}$.
These operations have the following definitions:
\begin{align}
    \Ph[\theta]{k} &= \exp\Big\{\iota\theta\ket{k}\bra{k}\Big\}, \\
    \label{eqn:rhpi}\Rp[\theta,\phi]{ij} &= \exp\left\{ -\iota\frac{\theta}{2} \sigma^{ij}_\phi \right\},
\end{align}
where $\sigma^{ij}_{\bullet}$ is the qudit analogue of qubit Pauli operators acting on the two-level subspace $\left\{\ket{i},\ket{j}\right\}$ of a qudit:
\begin{gather}
    \sigma^{ij}_x = \ket{j}\bra{i} + \ket{i}\bra{j}, \\
    \sigma^{ij}_y = \iota \ket{j}\bra{i} - \iota \ket{i}\bra{j}, \\
    \sigma^{ij}_\phi = e^{+\iota\phi} \ket{j}\bra{i} + e^{-\iota\phi} \ket{i}\bra{j}.
\end{gather}
Thus, $\Rp{ij}$ operation can be viewed as a $2{\times}2$ unitary matrix $\Rp{}$ that acts on subspace spanned by $\left\{\ket{i},\ket{j}\right\}$:
\begin{equation}
    \Rp[\theta, \phi]{} = \Matrix{\cos\left(\frac{\theta}{2}\right) & -\iota e^{-\iota\phi}\sin\left(\frac{\theta}{2}\right) \\ -\iota e^{+\iota\phi}\sin\left(\frac{\theta}{2}\right) & \cos\left(\frac{\theta}{2}\right)}.
\end{equation}

Several platforms may lack two-parametric transition operations, e.g., for photonic circuits, there is only a single parameter $r$ in the beam-splitter operation:
\begin{equation}
    \BS[r]{} = \Matrix{ r & -\sqrt{1-r^2} \\ \sqrt{1-r^2} & \sqrt{r} }.
\end{equation}
We may always apply phase shifts and transform this operation into an $\Rp{ij}$ operation:
\begin{equation}
    \Rp[\theta,\phi]{ij} = \Ph[\phi{-}\frac{\pi}{2}]{j}\circ\BS[\cos\frac{\theta}{2}]{ij}\circ\Ph[\frac{\pi}{2}{-}\phi]{j}.
\end{equation}

We also consider arbitrary selection rules for a given qudit physical implementation that permit applicable $\Rp{ij}$ operations.
In the following sections, we demonstrate that any qudit operation $\Utry{}$ can be decomposed into the sequence of allowed $\Rp{ij}$ and $\Ph{k}$ operations.
Moreover, for any selection rules, we guarantee at most $d(d{-} 1)/2$ transition operations $\Rp{ij}$ in the final decomposition.

\section{Unitary Decomposition for Trapped-Ion and Superconducting Hardware}
\label{section:decomposition-trapped-ion}

Existing decomposition schemes utilize superconducting selection rules: transitions are allowed only for neighboring states.
Two distinct schemes are provided in~\cite{Younis2023} (\textit{Row-by-row} scheme for superconducting circuits) and in~\cite{Clements2016} (\textit{Square} scheme for photonic circuits).
Both schemes perform unitary decomposition by eliminating non-diagonal entries from the unitary matrix $\Utry$ using a sequence of two-level $\Rp{}$ operations.

In these decompositions, we initialize the algorithm with a unitary $d{\times}d$ matrix.
After right-applying a two-level operation $\Rp[-\theta_1,\phi_1]{01}$ to $\Utry{}$,
elements of the first two columns are modified in the following way:
\begin{equation}
    \Matrix{\UtryElt{k}{0}' \\ \UtryElt{k}{1}'}
    = \Rp[-\theta_1, \phi_1]{}^T
    \Matrix{\UtryElt{k}{0} \\ \UtryElt{k}{1}}.
\end{equation}
By choosing proper values for parameters $\theta_1$ and $\phi_1$, we can set a single element of the first column $\mathcal{U}'_{d-1,0}$ to $0$:
\begin{gather}
\label{eqn:transition-params}
    \cos\left(\frac{\theta_1}{2}\right) \UtryElt{d-1}{0} + \iota e^{\iota\phi_1} \sin\left(\frac{\theta_1}{2}\right) \UtryElt{d-1}{1} = 0, \\
    \theta_1 = 2\arctan\left( \frac{|\UtryElt{d-1}{0}|}{|\UtryElt{d-1}{1}|} \right), \\
    \phi_1 = \frac{\pi}{2} + \arg(\UtryElt{d-1}{0}) - \arg(\UtryElt{d-1}{1}).
\end{gather}

In the next steps, we apply operations $\Rp{{n-1},{n}}$ in order to eliminate the element $\UtryElt{d-1}{n-1}$ using element $\UtryElt{d-1}{n}$.
After $d{-}1$ steps, all elements in a row $d{-}1$ are eliminated.
Due to the unitary property, all elements in a column $d{-}1$ are also eliminated, and the diagonal element has absolute value $1$.
With the phase gate $\Ph[\alpha_{d-1} = \arg(\UtryElt{d-1}{d-1})]{d-1}$, we obtain the following decomposition:
\begin{equation}
\begin{aligned}
\label{eqn:qudit-decomposition}
    \Utry[d] =~ & \Matrix{\Utry[d-1] & 0 \\ 0 & 1} \circ \\
    \circ~ & \Ph[\alpha_{d-1}]{d-1}
    \circ \Rp[\theta_{d-1},\phi_{d-1}]{d-2,d-1} \circ \\
    \circ~ & \cdots ~\circ \\
    \circ~ & \Rp[\theta_{2},\phi_{2}]{1,2} \circ \Rp[\theta_{1},\phi_{1}]{0,1}.
\end{aligned}
\end{equation}

\begin{figure*}[t]
    \begin{subfigure}{.47\linewidth}
        \includegraphics[width=\linewidth]{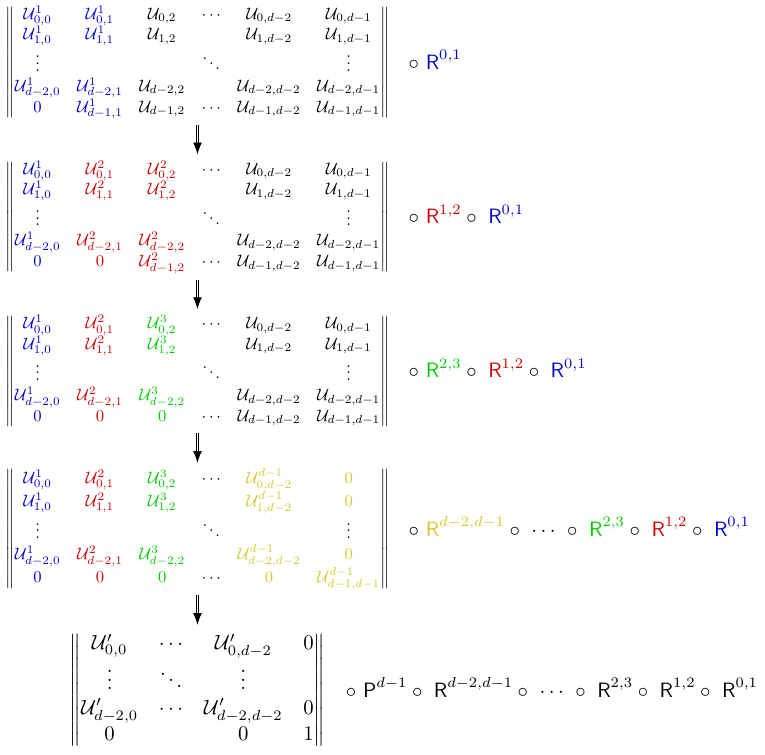}
        \caption{}
        \label{fig:decomp-scheme-line}
    \end{subfigure}
    \hfill
    \begin{subfigure}{.47\linewidth}
        \includegraphics[width=\linewidth]{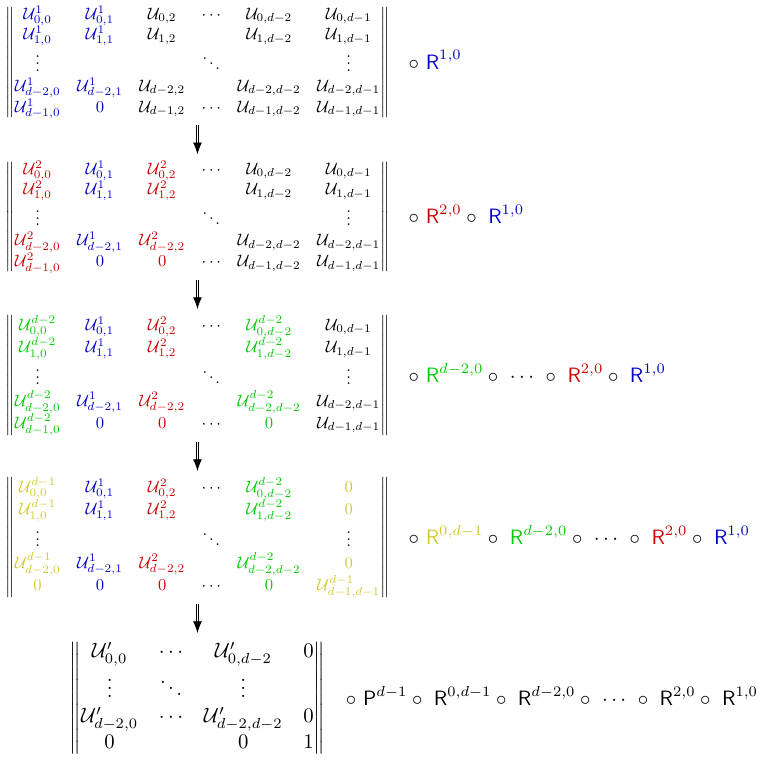}
        \caption{}
        \label{fig:decomp-scheme-star}
    \end{subfigure}

    \caption{
        Row elimination schemes for a qudit unitary matrix with given allowed transitions.
        (\textbf{a}) Decomposition for a superconducting qudit with allowed transitions $\Rp{n,{n \pm 1}}$.
        (\textbf{b}) Decomposition for a trapped-ion qudit with allowed transitions $\Rp{0,n}$ and $\Rp{n,0}$.}
    \label{fig:decomp-scheme}
\end{figure*}

After eliminating the row, we can repeat the same algorithm on the lesser $(d{-}1){\times}(d{-}1)$ unitary matrix $\Utry[d-1]$.
The full row elimination scheme for a superconducting qudit is depicted in \Cref{fig:decomp-scheme-line}.
Overall, this scheme requires at most $d(d{-}1)/2$ two-level transitions.

If we want to implement the similar algorithm on the trapped-ion quantum computer~\cite{Zalivako2024qb16} with $d{=}4$ levels, we face the following issue.
Consider the device operating with transitions $01$, $02$, $03$, whereas the given decomposition scheme operates with $01$, $12$, $23$.
The naive approach requires using level-swap operations that transform any transition $\Rp{ij}$ into the allowed transition $\Rp{0i}$:
\begin{equation}
\begin{aligned}
    \Rp[\theta,\phi]{ij} =&~\Sw{0i} \circ \Rp[\theta,\phi]{0j} \circ \Sw{i0}, \\
    &\text{or} \\
    \Rp[\theta,\phi]{ij} =&~\Sw{0j} \circ \Rp[\theta,\phi]{i0} \circ \Sw{j0},
\end{aligned}
\end{equation}
\begin{equation}
    \Sw{ij} = \Rp[\pi, \frac{\pi}{2}]{ij} \equiv \Matrix{0 & -1 \\ 1 & 0}.
\end{equation}
However, using this strategy for arbitrary unitary, we obtain $12$ trapped-ion native transitions, instead of $6$ as in the case of superconducting qudit.
The best we can get using level-swaps is $8$ native transitions.
This number will grow more rapidly than the estimation $\le d(d{-}1)/2$ for a superconducting qudit.

Hence, we should deduce the most optimal decomposition scheme for trapped-ion qudits, which typically have a star-like level transition graph.
In general, we consider $d$ levels trapped-ion qudit with transitions $\Rp{0,n}$ or $\Rp{n,0}$, where $0<n<d$.
The main idea of our algorithm is to eliminate all row elements $\UtryElt{d-1}{n}$ where $0<n<d$ using the zeroeth element $\UtryElt{d-1}{0}$.
Then, apply $\Rp{0,d-1}$ to eliminate the zeroeth element as well.
The full row elimination scheme for a trapped-ion qudit unitary is depicted in \Cref{fig:decomp-scheme-star}.

Row elimination for $d{\times}d$ unitary requires at most $(d{-}1){\times}~\Rp{}$ and $1{\times}~\Ph{}$ gates.
In total, after decomposing all lesser matrices, we obtain $d(d{-}1)/2{\times}~\Rp{}$ and $d{\times}~\Ph{}$ gates in unitary decomposition.
Though some $\theta_n$ parameters could be equal $0$ (if eliminated elements are already $=0$), hence these operations can be omitted in the hardware execution.
Also, on both considered platforms, phase gates $\Ph{}$ can be performed virtually, so their amount in the decomposition does not affect execution time as much as transition gates $\Rp{}$.

\section{Unitary Decomposition for Qudit with Arbitrary Selection Rules}
\label{section:decomposition-arbitrary-qudit}

Our goal is to generalize the considered decomposition schemes to the case of arbitrary selection rules.
A specific decomposition pattern could be spotted in the earlier proposed schemes.
We apply $d{-}1$ operations in order to eliminate all non-diagonal terms.
Each $\Rp{}$ operation requires an additional non-zero row element to perform elimination.
This pattern to eliminate a row $r$ could be described in a more formal way:
\begin{itemize}
    \item On each step, we apply $\Rp{z_ip_i}$ gate, where $z_i$ is the index of eliminated element in row and $p_i$ is the index of pivot element that used to eliminate $z_i$;
    \item $z_i, p_i$ must be allowed transition;
    \item Indices $z_i$ are non-repetitive and take all levels $\{0,1,\dots,r{-}1\}$;
    \item Indices $p_i$ could be arbitrary levels, except those that have already been eliminated:
    \begin{equation}
        \forall i<r,j<i \Rightarrow p_i \ne z_j.
    \end{equation}
\end{itemize}

In fact, we are not restricted to eliminating only the last row in the unitary matrix.
For a given sequence for eliminated rows $\{r_n\}_{n=0}^{d-1}$ (elimination from top $r_{d-1}$ to bottom $r_0$), we just have to modify the $3^\text{rd}$ rule in the pattern.
On $k^\text{th}$ step, where row $r_k$ is eliminated, we should run through indices $z_i$ that are non-repetitive and take all levels $\{r_0,r_1,\dots,r_{k-1}\}$;

In total, the decomposition scheme consists of indices:
\begin{itemize}
    \item $r_k$---the order of eliminated rows;
    \item $z_{ki}$---the order of eliminated elements on $k^\text{th}$ step;
    \item $p_{ki}$---pivot elements on $k^\text{th}$ step.
\end{itemize}
Using defined rules, we could produce a decomposition scheme for all arbitrary selection rules.
The existence of such decomposition is not obvious yet but will be proved during the rest of the paper.
However, indices for some selection rules, e.g., for a superconducting qudit and trapped-ion qudit, are already obtained in the previous section and depicted in \Cref{table:scheme-examples}.

\begin{table}[H]
    \centering

    \resizebox{\linewidth}{!}{\begin{tabular}{c|c|c|c|}
        \textbf{Platform} & ~\textbf{d}~ & \textbf{Connectivity} & \textbf{Decomposition Schemes} \\
        \hline
        Superconducting qudit & 4 &
        \raisebox{-0.45\totalheight}{\includegraphics[height=2cm]{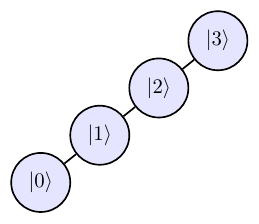}}
        & $\begin{matrix} r_3 = 3 & z_3 = \{ 0, 1, 2 \} & p_3 = \{ 1, 2, 3 \} \\ r_2 = 2 & z_2 = \{ 0, 1 \} & p_2 = \{ 1, 2 \} \\ r_1 = 1 & z_1 = \{ 0 \} & p_1 = \{ 1 \} \\ \end{matrix}$ \\
        \hline
        $^{171}\text{Yb}^+$ trapped-ion qudit & 4 &
        \raisebox{-0.45\totalheight}{\includegraphics[height=2cm]{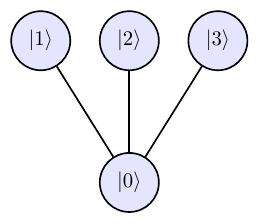}}
        & $\begin{matrix} r_3 = 3 & z_3 = \{ 1, 2, 0 \} & p_3 = \{ 0, 0, 3 \} \\ r_2 = 2 & z_2 = \{ 1, 0 \} & p_2 = \{ 0, 2 \} \\ r_1 = 1 & z_1 = \{ 0 \} & p_1 = \{ 1 \} \\ \end{matrix}$ \\
        \hline
        $^{171}\text{Yb}^+$ trapped-ion qudit & 5 &
        \raisebox{-0.45\totalheight}{\includegraphics[height=2cm]{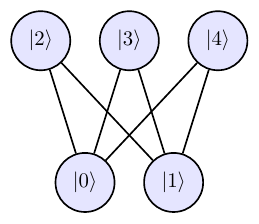}}
        & $\begin{matrix} r_4 = 1 & z_4 = \{ 0, 4, 3, 2 \} & p_4 = \{ 2, 1, 1, 1 \} \\ r_3 = 4 & z_3 = \{ 3, 2, 0 \} & p_3 = \{ 0, 0, 4 \} \\ r_2 = 3 & z_2 = \{ 2, 0 \} & p_2 = \{ 0, 3 \} \\ r_1 = 2 & z_1 = \{ 0 \} & p_1 = \{ 2 \} \\ \\ r_4 = 4 & z_4 = \{ 3, 2, 1, 0 \} & p_4 = \{ 0, 0, 4, 4 \} \\ r_3 = 3 & z_3 = \{ 2, 1, 0 \} & p_3 = \{ 0, 3, 3 \} \\ r_2 = 1 & z_2 = \{ 0, 2 \} & p_2 = \{ 2, 1 \} \\ r_1 = 2 & z_1 = \{ 0 \} & p_1 = \{ 2 \} \\ \end{matrix}$ \\
        \hline
    \end{tabular}}

    \caption{Decomposition scheme indices for several qudit platforms: superconducting with 4 levels and trapped-ion with 4 and 5 levels.}
    \label{table:scheme-examples}
\end{table}

We note that these rules are not sufficient to get a unique decomposition scheme.
For example, in the trapped-ion case, we can eliminate the first $k{-}1$ elements in any order and still get the row eliminated.
Also, a trapped-ion qudit with $5$ levels can utilize two different schemes with different row orders.
This degree of freedom can be useful, since we are allowed to select a preferable sequence of operations according to physical heuristics of levels and transitions of the given qudit platform.

An arbitrary single-qudit system could be described as a connected undirected graph $\mathcal{G}$, where nodes $\mathcal{N}$ represent qudit levels and edges $\mathcal{E}$ represent allowed transitions.
To determine decomposition indices for a given graph $\mathcal{G}$, we propose the following algorithm:

\begin{figure}[t!]
    \begin{subfigure}{\linewidth}
        \centering
        \includegraphics[width=.7\linewidth]{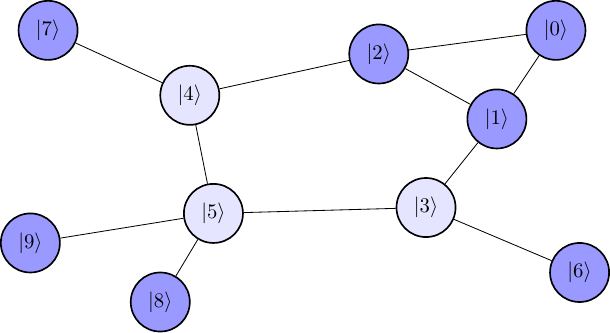}
        \caption{Select $\ket{r}$ candidates that can be removed without breaking connectivity of the graph.}
        \label{fig:trans-graph-a}
    \end{subfigure}

    \begin{subfigure}{\linewidth}
        \centering
        \includegraphics[width=.7\linewidth]{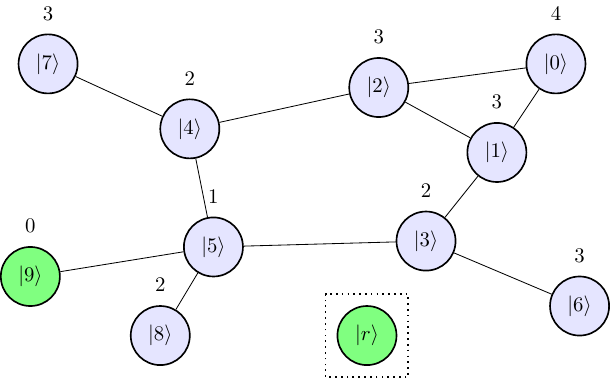}
        \caption{Split the transition graph into sets $\{F_l\}$ using breadth-first search.
        $\forall l,\ket{n}\in F_l \to \text{dist}(\ket{r},\ket{n}) = l$.}
        \label{fig:trans-graph-b}
    \end{subfigure}

    \begin{subfigure}{\linewidth}
        \centering
        \includegraphics[width=.7\linewidth]{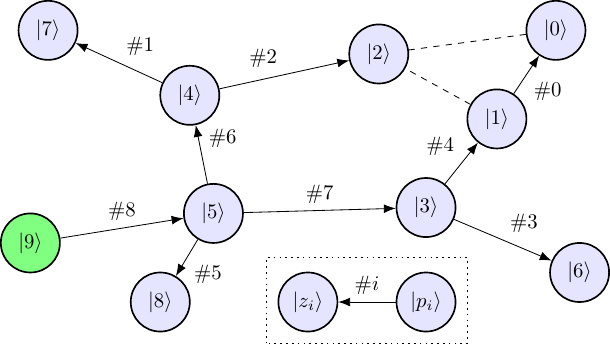}
        \caption{Add indices $\ket{z_i}$ and $\ket{p_i}$ into the decomposition scheme for row $\ket{r}$.}
        \label{fig:trans-graph-c}
    \end{subfigure}

    \begin{subfigure}{\linewidth}
        \centering
        \includegraphics[width=.7\linewidth]{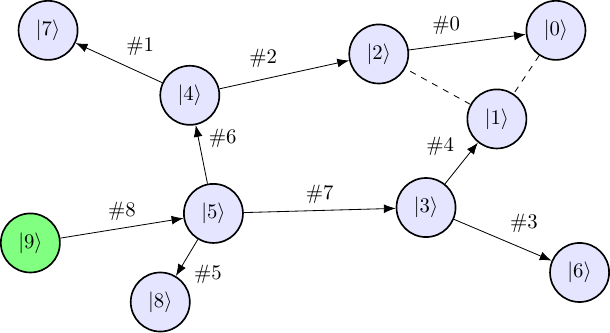}
        \caption{Degree of freedom is presented in this algorithm.
        Qudit physical properties could be used to choose the best transition sequence.}
        \label{fig:trans-graph-d}
    \end{subfigure}

    \caption{Algorithm that determines indices for the qudit decomposition scheme.
    It utilizes the graph structure of selection rules and the breadth-first search algorithm.}
    \label{fig:trans-graph}
\end{figure}

\begin{itemize}
    \item On the step $k$ (runs from $d{-}1$ to $1$), $\mathcal{N}$ has $k{+}1$ nodes/levels.
    We select levels that, if removed, do not break connectivity of the graph.
    One of these levels can be chosen as $\ket{r_k}$ to eliminate the row $r_k$ of the qudit unitary matrix (\Cref{fig:trans-graph-a}).
    Here, we can use physical heuristics to eliminate less stable levels earlier.
    \item Each level $\ket{i}\in\mathcal{G}$ is assigned the shortest distance from the level $\ket{r_k}$.
    This could be done using breadth-first search, which has time complexity $O(\left|\mathcal{N}\right| + \left|\mathcal{E}\right|)$.
    We can split all levels into sets $\{F_l\}$, where $F_l$ contains levels at distance $l$ from $\ket{r_k}$ (\Cref{fig:trans-graph-b}).
    \item Let $L$ be the maximum distance from $\ket{r_k}$ in the $\mathcal{G}$.
    Each level $\ket{z_{L,i}} \in F_L$ is connected by a transition to some element $\ket{p_{L,i}} \in F_{L-1}$.
    We add these indices into the decomposition scheme, then repeat for all $F_l$ in descending order of $l$ (\Cref{fig:trans-graph-c}).
    There could be several levels $\in F_{L-1}$ connected to $\ket{z_{L,i}}$ (\Cref{fig:trans-graph-d}).
    Using physical properties of a qudit, we can select the least noisy transition to reduce the average error of hardware execution.
    \item Continue the algorithm on the step $k{-}1$ for the transition graph $\mathcal{G}' = \mathcal{G} / \{\ket{r_k}\}$ until $k=0$.
\end{itemize}
Produced decomposition indices satisfy rules required for correct row elimination:
\begin{itemize}
    \item Indices $z_{l,i}$ are taken from each $\{F_l\}_{l=1}^{L}$. These are non-intersecting sets, so $\{\ket{z_i}\} = \mathcal{G} / \{\ket{r_k}\}$, i.e., all elements are eliminated except the diagonal one.
    \item Indices $p_{l,i}$ do not intersect already eliminated elements, since they have distance $>l$.
\end{itemize}

We are able to estimate the overall time complexity of proposed algorithm.
Each step has complexity $O(\left|\mathcal{N}\right| + \left|\mathcal{E}\right|)$ from the breadth-first search and $O(\left|\mathcal{N}\right|)$ from assigning physical rotations.
Thus, they are summed up to $O(d^2 + Td)$ complexity, where $T$ is the total number of allowed transitions within the qudit.

Note that on each step transition graph remains connected; this property is crucial for correct unitary decomposition.
If we have a graph with two or more disconnected components, then we can only obtain a non-zero element per component.
Each of these non-zero elements could not be eliminated, since they are not connected via any $\Rp{}$ path.

The similar graph algorithm for the transition graph of the $^{86}\text{Rb}$ atom qudit was mentioned in~\cite{Bullock2005}.
Though it describes the general technique, we propose an algorithm that is able to minimize physical execution error and reduce the number of transitions for sparse unitary matrices.

\begin{figure}[t!]
    \begin{subfigure}{\linewidth}
        \centering
        \includegraphics[height=.5\linewidth]{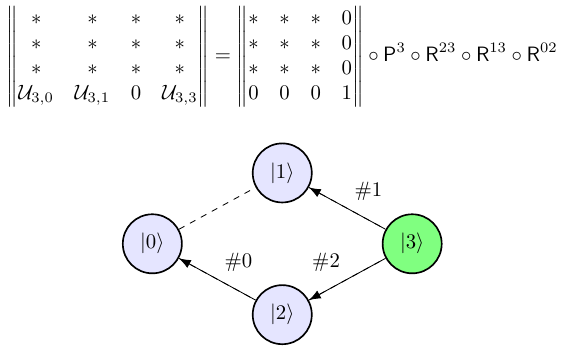}
        \caption{}
        \label{fig:trans-graph-mode-static}
    \end{subfigure}

    \vspace{.5cm}

    \begin{subfigure}{\linewidth}
        \centering
        \includegraphics[height=.5\linewidth]{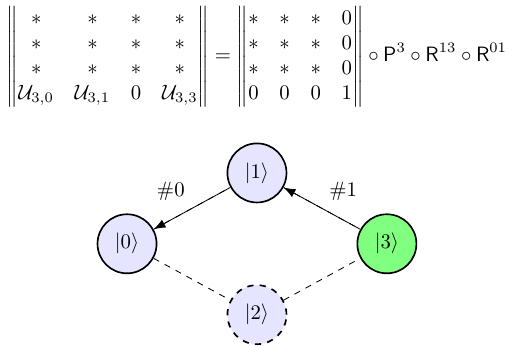}
        \caption{}
        \label{fig:trans-graph-mode-adaptive}
    \end{subfigure}

    \caption{Decomposition of a matrix with zero non-diagonal elements: (\textbf{a}) \textit{Static} and (\textbf{b}) \textit{Adaptive} schemes.}
    \label{fig:trans-graph-mode}
\end{figure}

The \textit{static} decomposition scheme is obtained once for a given qudit platform.
Then, this scheme can be used for any qudit unitary matrix that should be executed on a given qudit.
However, for some sparse matrices with zero non-diagonal elements, better-optimized decomposition may be performed using the \textit{adaptive} decomposition scheme (\Cref{fig:trans-graph-mode}).
The adaptive decomposition runs through the same step as static decomposition for each unitary matrix.
The difference is that a level is excluded from the graph during decomposition if it corresponds to a zero element and does not break graph connectivity.
Also, the order of eliminated rows is determined by the matrix sparsity pattern: rows with fewer non-zero elements are eliminated earlier.
To sum up, the \textit{adaptive} decomposition scheme is only valid for a given matrix and contains the least number of operations, whereas the \textit{static} decomposition scheme can be applied to an arbitrary matrix.

\section{Comparison with Existing Decomposition Methods}
\label{section:comparison}

To show the advantage of the developed algorithm, we provide the comparison with other approaches that provide the functionality of decomposing arbitrary qudit unitary operations into the sequence of native pulses.
All of them provide Python packages that help execute and compare results on a $d{\times}d$ Python matrix of the type \texttt{np.ndarray}.
We compare existing methods with our developed decompositions in \Cref{table:comparison-counts,table:comparison-times}.
All experiments were conducted using a MacBook Pro 2020 with an Apple M1 chip.

\begin{table*}[t]
    \centering

    \begin{subtable}{\textwidth}
        \centering
        \begin{tabular}{l|ccc|ccc|ccc|ccc|cccccc|}
            \multirow{2}{*}{\textbf{Method}}
            & \multicolumn{3}{c|}{$\Gate{X}{}{d}{+1}$} & \multicolumn{3}{c|}{$\Gate{X}{}{d}{-1}$} & \multicolumn{3}{c|}{$\Gate{H}{}{d}{}$}
            & \multicolumn{3}{c|}{$\Utry[d]$} & \multicolumn{6}{c|}{\textbf{Two-Qubit Gate}} \\
            & \boldmath{$d{=}4$} & \boldmath{$d{=}5$} & \boldmath{$d{=}6$} & \boldmath{$d{=}4$} & \boldmath{$d{=}5$} & \boldmath{$d{=}6$} & \boldmath{$d{=}4$} & \boldmath{$d{=}5$} & \boldmath{$d{=}6$}
            & \boldmath{$d{=}4$} & \boldmath{$d{=}5$} & \boldmath{$d{=}6$} & $\Gate{RXX}{}{}{}$ & $\Gate{RZZ}{}{}{}$ & $\Gate{CZ}{}{}{}$ & $\Gate{CX}{}{}{}$ & $\Gate{CH}{}{}{}$ & $\Gate{SWAP}{}{}{}$ \\
            \hline
            QSearchPass~\cite{Davis2020}
            & 3     & 4     & 5     & 3     & 4     & 5     & 7     & 11    & 17
            & 7     & 12    & 16    & 6     & 0     & 0     & 3     & 3     & 1     \\
            QSweepPass~\cite{Younis2023}
            & 5     & 6     & 7     & 3     & 4     & 5     & 7     & 11    & 16
            & 7     & 11    & 16    & 6     & 2     & 2     & 5     & 5     & 3     \\
            LocQRPass~\cite{Mato2024}
            & 3     & 4     & 5     & 3     & 4     & 5     & 6     & 10    & 15
            & 6     & 10    & 15    & 6     & 0     & 0     & 3     & 3     & 1     \\
            LocAdaPass~\cite{Mato2024}
            & 3     & 4     & 5     & 3     & 4     & 5     & 6     & \TO{} & \TO{}
            & 7     & \TO{} & \TO{} & 4     & 0     & 0     & 2     & 2     & 1     \\
            \hline
            TAQR
            & 3     & 4     & 5     & 3     & 4     & 5     & 6     & 10    & 15
            & 6     & 10    & 15    & 6     & 0     & 0     & 3     & 3     & 1     \\
            TAQR Adaptive
            & 3     & 4     & 5     & 3     & 4     & 5     & 6     & 10    & 15
            & 6     & 10    & 15    & 6     & 0     & 0     & 3     & 3     & 1     \\
            \hline
        \end{tabular}
        \caption{Line transition graph.}
        \label{table:comparison-counts-line}
    \end{subtable}

    \vspace{.1cm}

    \begin{subtable}{\textwidth}
        \centering
        \begin{tabular}{l|ccc|ccc|ccc|ccc|cccccc|}
            \multirow{2}{*}{\textbf{Method}}
            & \multicolumn{3}{c|}{$\Gate{X}{}{d}{+1}$} & \multicolumn{3}{c|}{$\Gate{X}{}{d}{-1}$} & \multicolumn{3}{c|}{$\Gate{H}{}{d}{}$}
            & \multicolumn{3}{c|}{$\Utry[d]$} & \multicolumn{6}{c|}{\textbf{Two-Qubit Gate}} \\
            & \boldmath{$d{=}4$} & \boldmath{$d{=}5$} & \boldmath{$d{=}6$} & \boldmath{$d{=}4$} & \boldmath{$d{=}5$} & \boldmath{$d{=}6$} & \boldmath{$d{=}4$} & \boldmath{$d{=}5$} & \boldmath{$d{=}6$}
            & \boldmath{$d{=}4$} & \boldmath{$d{=}5$} & \boldmath{$d{=}6$} & $\Gate{RXX}{}{}{}$ & $\Gate{RZZ}{}{}{}$ & $\Gate{CZ}{}{}{}$ & $\Gate{CX}{}{}{}$ & $\Gate{CH}{}{}{}$ & $\Gate{SWAP}{}{}{}$ \\
            \hline
            QSearchPass~\cite{Davis2020}
            & 3     & 4     & 5     & 3     & 4     & 5     & 6     & 12    & 16
            & 7     & 11    & 16    & 4     & 0     & 0     & 3     & 3     & 3     \\
            LocQRPass~\cite{Mato2024}
            & 7     & 10    & 13    & 7     & 10    & 13    & 16    & 28    & 43
            & 16    & 28    & 43    & 16    & 0     & 0     & 9     & 9     & 3     \\
            LocAdaPass~\cite{Mato2024}
            & 4     & 6     & 7     & 4     & 5     & 6     & 8     & 15    & 24
            & 8     & 15    & 24    & 3     & 0     & 0     & 2     & 2     & 2     \\
            \hline
            TAQR
            & 5     & 6     & 9     & 3     & 4     & 5     & 5     & 10    & 15
            & 6     & 10    & 15    & 4     & 0     & 0     & 3     & 3     & 3     \\
            TAQR Adaptive
            & 3     & 4     & 5     & 3     & 4     & 5     & 6     & 10    & 15
            & 6     & 10    & 15    & 4     & 0     & 0     & 3     & 3     & 3     \\
            \hline
        \end{tabular}
        \caption{Star transition graph.}
        \label{table:comparison-counts-star}
    \end{subtable}

    \vspace{.1cm}

    \begin{subtable}{\textwidth}
        \centering
        \begin{tabular}{l|ccc|ccc|ccc|ccc|cccccc|}
            \multirow{2}{*}{\textbf{Method}}
            & \multicolumn{3}{c|}{$\Gate{X}{}{d}{+1}$} & \multicolumn{3}{c|}{$\Gate{X}{}{d}{-1}$} & \multicolumn{3}{c|}{$\Gate{H}{}{d}{}$}
            & \multicolumn{3}{c|}{$\Utry[d]$} & \multicolumn{6}{c|}{\textbf{Two-Qubit Gate}} \\
            & \boldmath{$d{=}4$} & \boldmath{$d{=}5$} & \boldmath{$d{=}6$} & \boldmath{$d{=}4$} & \boldmath{$d{=}5$} & \boldmath{$d{=}6$} & \boldmath{$d{=}4$} & \boldmath{$d{=}5$} & \boldmath{$d{=}6$}
            & \boldmath{$d{=}4$} & \boldmath{$d{=}5$} & \boldmath{$d{=}6$} & $\Gate{RXX}{}{}{}$ & $\Gate{RZZ}{}{}{}$ & $\Gate{CZ}{}{}{}$ & $\Gate{CX}{}{}{}$ & $\Gate{CH}{}{}{}$ & $\Gate{SWAP}{}{}{}$ \\
            \hline
            QSearchPass~\cite{Davis2020}
            & 3     & 4     & 6     & 3     & 4     & 5     & 7     & 12    & 13
            & 7     & 11    & 16    & 2     & 0     & 0     & 1     & 1     & 1     \\
            LocQRPass~\cite{Mato2024}
            & 7     & 10    & 13    & 7     & 10    & 13    & 14    & 26    & 41
            & 14    & 26    & 41    & 14    & 0     & 0     & 7     & 7     & 1     \\
            LocAdaPass~\cite{Mato2024}
            & 4     & 6     & 7     & 4     & 5     & 6     & 9     & 14    & 21
            & 9     & 14    & 21    & 2     & 0     & 0     & 1     & 1     & 1     \\
            \hline
            TAQR
            & 3     & 6     & 7     & 3     & 4     & 5     & 6     & 10    & 14
            & 6     & 10    & 15    & 2     & 0     & 0     & 1     & 1     & 1     \\
            TAQR Adaptive
            & 3     & 4     & 5     & 3     & 4     & 5     & 6     & 10    & 14
            & 6     & 10    & 15    & 2     & 0     & 0     & 1     & 1     & 1     \\
            \hline
        \end{tabular}
        \caption{Bipartite transition graph.}
        \label{table:comparison-counts-bipartite}
    \end{subtable}

    \caption{Comparison of decompositions produced by the given methods in terms of transition count.
    \TO{} means that the execution exceeded the 1 min timeout.}
    \label{table:comparison-counts}
\end{table*}

The first approach is a \textit{BQSKIT} framework that provides a generic method for numerical synthesis of unitary operations.
It supports a qudit unitary synthesis for multiqubit and multiqudit circuits.
There exist several methods in \textit{BQSKIT} that allow one to decompose qudit unitary operations.
First is the \textit{QSearchPass}~\cite{Davis2020} method, whose main purpose is to synthesize multiqubit circuits using the native gate set according to qubit connectivity, though it is not limited only to qubits processing and can be utilized to synthesize single-qudit unitary operations.
We can obtain desired decomposition by declaring custom parametric qudit gates, which correspond to allowed transition $\Rp{ij}$ and phase shift $\Ph{k}$ operations.
Notably, this method is not deterministic, so we run each experiment 20 times and take the median value for decomposition length.
\Cref{listing:code-qsearch} contains Python code to invoke \textit{QSearchPass}.
The second method is called \textit{QSweepPass}~\cite{Younis2023}, which is not part of \textit{BQSKIT} but relies on its transpilation workflow.
It uses numerical and analytical results to construct a qudit unitary operation in terms of two-level operations.
However, it does not support arbitrary selection rules and only targets superconducting transition topology.
\Cref{listing:code-qsweep} contains Python code to invoke \textit{QSweepPass}.

The second approach is provided by \textit{MQT.Qudits}~\cite{Mato2024} framework for mixed-dimensional quantum computing, which is part of the \textit{Munich Quantum Toolkit}.
It is possible to pass a qudit transition graph with corresponding weights into the \textit{LocQRPass} method and obtain the sequence of allowed gates as decomposition.
This framework also supports the adaptive gate transpilation mode via the \textit{LocAdaPass} method that can reduce the length of the decomposition.
\Cref{listing:code-mqt} contains Python code to invoke \textit{MQT.Qudits}.

In \Cref{table:comparison-counts}, our method is called \textit{Transition-Aware QR decomposition}, or \textit{TAQR} for simplicity, for static decomposition and \textit{TAQR Adaptive} for adaptive decomposition.
In our comparison, we are targeting three types of selection rules given as an undirected graph for different qudit platforms.
The first considered platform has \textit{the line transition graph} (\Cref{table:comparison-counts-line}), where only transitions on neighboring levels are allowed.
Superconducting qudits and photonic circuits have this type of selection rules.
We can extend the line transition graph to any number of levels $d$.
The second platform consists of trapped-ion qudits.
Overall, this type of qudit can have any arbitrary transitions selected by atomic physics rules.
Though, particular trapped-ion devices provide \textit{the star transition graph} (\Cref{table:comparison-counts-star}) or \textit{the bipartite transition graph} (\Cref{table:comparison-counts-bipartite}) of the size $d$.
\textit{The star transition graph} is the special case of \textit{the bipartite transition graph} with partition $p{=}1$.
\textit{The bipartite transition graph} with a partition $p$ is the graph with two disjoint subsets with sizes $p$ and $d{-}p$, that is, every node in the first subset is connected via an edge to each node in the second subset.
We use bipartite graph with $p{=}2$ in our experiments.

Each method produces decomposition for each given platform for one of the following single-qudit operations:
\begin{itemize}
    \item State number increment/decrement gate $\Gate{X}{}{d}{+1}$ and $\Gate{X}{}{d}{-1}$, which have the form as in \Cref{eqn:qudit-x-gate}.
    \item Quantum Fourier Transform with a size $d$ that is denoted $\Gate{H}{}{d}{}$ in \Cref{eqn:qudit-h-gate}.
    \item Uniformly distributed (using Haar measure procedure) unitary matrix $\Utry[d]$. Since the matrix is random, we run each decomposition on 100 generated matrices, then take median value for decomposition length.
    \item Two-qubit matrix $\in U(4)$ that represents the action on two qubits embedded into the qudit with $d{=}4$.
    Chosen gates are as follows: $\Gate{RXX}{\pi/2}{}{}$, $\Gate{RZZ}{\pi/2}{}{}$, $\Gate{CZ}{}{}{}$, $\Gate{CX}{}{}{}$, $\Gate{CH}{}{}{}$, and $\Gate{SWAP}{}{}{}$.
\end{itemize}

As a result, \textit{Transition-Aware QR} decomposition performs as good as any other qudit synthesis framework for superconducting or photonic platforms.
For the majority of qudit unitary operations, it produces the decomposition with the same length in a reasonable time (refer \Cref{table:comparison-times} in the \Cref{app:a} for execution time comparison).
Thus, it can become a drop-in replacement for the qudit transpiler pass for single-qudit operations for this type of platform.

Next, our method provides the shortest decomposition for trapped-ion platforms (with any arbitrary selection rules).
Our result outperforms the average numerical decomposition using \textit{QSearch} method that produces random decomposition on each invocation and takes significantly greater execution time.
It also outperforms \textit{MQT.Qudits} compilation passes, which produce more transitions and level-swaps for the same unitary.
For generic unitary, our method is guaranteed to produce $d(d{-}1)/2$ transitions, which is the theoretical upper bound (the number of lower non-diagonal elements in unitary).

Furthermore, lastly, we have shown the potential benefits of using either \textit{static} or \textit{adaptive} decomposition.
Static decomposition is computed once for each platform, which reduces decomposition time during the transpilation process.
On the other hand, adaptive decomposition considers the sparsity pattern of the given unitary matrix and produces a sequence with fewer transitions for several cases, e.g., for the $\Gate{X}{}{d}{}$ operation.

The total error of single-qudit execution grows with the number of transitions involved.
The error of single-qudit transition $\varepsilon_{ij}$ is quantified as an operator norm of a difference between the hardware implementation of $\Rp{ij}$ and its theoretical expression in \Cref{eqn:rhpi}.
Given the sequence of transitions $\mathcal{D}$, we can obtain the estimation for the total error $\varepsilon_{\rm tot}$ of unitary execution:
\begin{equation}
    \varepsilon_{\rm tot} \leq \sum_{ij \in \mathcal{D}} \varepsilon_{ij}.
\end{equation}
Hence, minimizing the number of transitions in decomposition leads to more accurate computations.
Since our approach produces shorter sequences for many cases, it also performs better in mitigation of execution errors on qudit hardware.
Moreover, the algorithm is able to choose transition with minimal error $\varepsilon_{ij}$ on the particular steps, if we are able to provide the estimation for transition error beforehand.

Notably, while \textit{LocAdaPass} produces shorter decomposition for two-qubit gates embedded into the ququart, it also permutes levels virtually.
This fact makes sense for optimization of the whole qudit circuit, though we should permute rows virtually or use level-swap gates to obtain proper unitary.
Potentially, our method can also benefit from virtual level-swaps when it is a part of the whole qudit circuit workflow.

\section{Conclusions}

In this paper, we have proposed a method that converts a $d$-level qudit unitary operation into the sequence of the allowed transitions between qudit levels.
It considers arbitrary selection rules of the qudit system and produces the optimal decomposition sequence in terms of the number of transitions.
Moreover, we show that the decomposition consists of no more than $d(d{-}1)/2$ transitions, which is the theoretical upper bound for an arbitrary unitary matrix $d{\times}d$.

Decomposition could be performed once for each qudit platform of interest before the transpilation, then applied to any unitary matrix to obtain parameters of transitions.
This is the essence of the \textit{static} decomposition.
On the other hand, we have proposed the \textit{adaptive} modification that utilizes the sparsity pattern of the matrix and the transition graph simultaneously.
Such treatment could reduce the number of transitions even more for some operations, such as level permutation.

As we have shown, the crucial advantage of our approach is that it does not distinguish between qudit platforms and produces similar optimal decomposition for any transition graph.
Most trapped-ion qudit platforms could benefit from utilizing this approach due to complex selection rules on operated levels.

Thus, the developed algorithm with the \textit{static} and \textit{adaptive} decompositions could become a drop-in replacement for single-qudit transpilation pass within qudit transpiler for multiqudit circuits.
Multiqudit operations have an analytic decomposition in terms of native two-qudit operations and single-qudit generic unitary operation.
TAQR could be utilized to decompose generic operations into an optimized sequence of native single-qudit operations and reduce single-qudit errors, therefore mitigating overall circuit execution error.

\section*{Author Contributions}

Conceptualization, A.S.N.;
Software, D.A.D.;
Validation, E.O.K.;
Investigation, D.A.D.;
Writing---original draft, D.A.D.;
Writing---review and editing, A.K.F., E.O.K. and A.S.N.;
Supervision, A.S.N. and A.K.F.;
Funding acquisition, A.S.N. and A.K.F.
All authors have read and agreed to the published version of the manuscript.

\section*{Funding}

The work of A.S.N. and D.A.D. was supported by RSF Grant No.~24-71-00084 (developing methods of quantum algorithms transpilation for the acceleration of their execution; \Cref{section:decomposition-trapped-ion,section:decomposition-arbitrary-qudit,,section:comparison}).
The work of E.O.K. and A.K.F. was supported by the Priority 2030 program at the NUST ``MISIS'' under the project K1-2022-027 (exploring existing qudit gates; \Cref{section:qudit-logic,section:qudit-hardware}).

\section*{Institutional Review Board Statement}

Not applicable.

\section*{Data Availability Statement}

The source code of the \textit{TAQR} comparison with other methods is available at \url{https://github.com/d-drozhzhin/qudit-unitary-decomposition}.
The source code of the developed method is available on reasonable request.

\section*{Conflicts of Interest}

The authors declare no conflicts of interest.

\bibliography{qudit-unitary.bib}

\newpage
\onecolumngrid
\appendix
\crefalias{section}{appendix}

\section{Comparison with Existing Decomposition Methods in Terms of Execution Time}
\label{app:a}

\begin{table}[H]
    \centering

    \begin{subtable}{\textwidth}
        \centering
        \begin{tabular}{l|ccc|ccc|ccc|}
            \multirow{2}{*}{\textbf{Method}}
            & \multicolumn{3}{c|}{$\Gate{X}{}{d}{+1}$} & \multicolumn{3}{c|}{$\Gate{X}{}{d}{-1}$} & \multicolumn{3}{c|}{$\Gate{H}{}{d}{}$} \\
            & \boldmath{$d{=}4$} & \boldmath{$d{=}5$} & \boldmath{$d{=}6$} & \boldmath{$d{=}4$} & \boldmath{$d{=}5$} & \boldmath{$d{=}6$} & \boldmath{$d{=}4$} & \boldmath{$d{=}5$} & \boldmath{$d{=}6$} \\
            \hline
            QSearchPass~\cite{Davis2020}
            & 3182      & 14078     & 10513     & 3160      & 7046      & 14449     & 3712      & 6264      & 16764     \\
            QSweepPass~\cite{Younis2023}
            & 61.95     & 96.84     & 157.01    & 13.02     & 17.37     & 24.71     & 145.45    & 246.6     & 427.24    \\
            LocQRPass~\cite{Mato2024}
            & 0.227     & 0.294     & 0.328     & 0.239     & 0.298     & 0.552     & 0.421     & 0.785     & 0.789     \\
            LocAdaPass~\cite{Mato2024}
            & 0.706     & 0.93      & 1.223     & 1.263     & 3.23      & 8.81      & 1427.2    & \TO{}     & \TO{}     \\
            \hline
            TAQR
            & 0.28      & 0.347     & 0.487     & 0.281     & 0.425     & 0.468     & 0.435     & 0.688     & 1.009     \\
            TAQR Adaptive
            & 1.261     & 2.318     & 4.456     & 1.253     & 2.341     & 4.146     & 1.553     & 2.854     & 4.878     \\
            \hline
        \end{tabular}
        \begin{tabular}{l|ccc|cccccc|}
            \multirow{2}{*}{\textbf{Method}}
            & \multicolumn{3}{c|}{$\Utry[d]$} & \multicolumn{6}{c|}{\textbf{Two-Qubit Gate}} \\
            & \boldmath{$d{=}4$} & \boldmath{$d{=}5$} & \boldmath{$d{=}6$} & $\Gate{RXX}{}{}{}$ & $\Gate{RZZ}{}{}{}$ & $\Gate{CZ}{}{}{}$ & $\Gate{CX}{}{}{}$ & $\Gate{CH}{}{}{}$ & $\Gate{SWAP}{}{}{}$ \\
            \hline
            QSearchPass~\cite{Davis2020}
            & 3993      & 7324      & 16364     & 10378     & 2554      & 2204      & 2891      & 3170      & 2550      \\
            QSweepPass~\cite{Younis2023}
            & 155.2     & 323.21    & 599.99    & 57.8      & 48.01     & 39.79     & 52.21     & 74.06     & 44.62     \\
            LocQRPass~\cite{Mato2024}
            & 0.396     & 0.553     & 0.998     & 0.361     & 0.097     & 0.083     & 0.229     & 0.237     & 0.147     \\
            LocAdaPass~\cite{Mato2024}
            & 1089.1    & \TO{}     & \TO{}     & 0.929     & 0.179     & 0.15      & 0.472     & 0.478     & 0.249     \\
            \hline
            TAQR
            & 0.494     & 0.692     & 1.035     & 0.407     & 0.117     & 0.105     & 0.265     & 0.282     & 0.166     \\
            TAQR Adaptive
            & 1.67      & 2.857     & 4.861     & 1.502     & 0.597     & 0.594     & 1.638     & 1.185     & 0.859     \\
            \hline
        \end{tabular}
        \caption{Line transition graph.}
        \label{table:comparison-times-line}
    \end{subtable}

    \vspace{.1cm}

    \begin{subtable}{\textwidth}
        \centering
        \begin{tabular}{l|ccc|ccc|ccc|}
            \multirow{2}{*}{\textbf{Method}}
            & \multicolumn{3}{c|}{$\Gate{X}{}{d}{+1}$} & \multicolumn{3}{c|}{$\Gate{X}{}{d}{-1}$} & \multicolumn{3}{c|}{$\Gate{H}{}{d}{}$} \\
            & \boldmath{$d{=}4$} & \boldmath{$d{=}5$} & \boldmath{$d{=}6$} & \boldmath{$d{=}4$} & \boldmath{$d{=}5$} & \boldmath{$d{=}6$} & \boldmath{$d{=}4$} & \boldmath{$d{=}5$} & \boldmath{$d{=}6$} \\
            \hline
            QSearchPass~\cite{Davis2020}
            & 2821      & 3067      & 3742      & 2929      & 3057      & 4241      & 4147      & 7187      & 17974     \\
            LocQRPass~\cite{Mato2024}
            & 0.392     & 0.896     & 0.625     & 0.361     & 0.514     & 0.654     & 0.842     & 1.361     & 2.036     \\
            LocAdaPass~\cite{Mato2024}
            & 1.052     & 2.222     & 2.545     & 1.005     & 1.509     & 2.144     & 4.606     & 14.07     & 36.08     \\
            \hline
            TAQR
            & 0.663     & 0.472     & 0.623     & 0.274     & 0.367     & 0.459     & 0.397     & 1.09      & 0.99      \\
            TAQR Adaptive
            & 1.359     & 2.204     & 2.374     & 1.059     & 1.631     & 2.355     & 1.379     & 2.335     & 3.578     \\
            \hline
        \end{tabular}
        \begin{tabular}{l|ccc|cccccc|}
            \multirow{2}{*}{\textbf{Method}}
            & \multicolumn{3}{c|}{$\Utry[d]$} & \multicolumn{6}{c|}{\textbf{Two-Qubit Gate}} \\
            & \boldmath{$d{=}4$} & \boldmath{$d{=}5$} & \boldmath{$d{=}6$} & $\Gate{RXX}{}{}{}$ & $\Gate{RZZ}{}{}{}$ & $\Gate{CZ}{}{}{}$ & $\Gate{CX}{}{}{}$ & $\Gate{CH}{}{}{}$ & $\Gate{SWAP}{}{}{}$ \\
            \hline
            QSearchPass~\cite{Davis2020}
            & 3949      & 7785      & 17719     & 3461      & 2602      & 2200      & 3068      & 3107      & 3051      \\
            LocQRPass~\cite{Mato2024}
            & 1.186     & 1.173     & 1.947     & 0.713     & 0.101     & 0.083     & 0.407     & 0.408     & 0.208     \\
            LocAdaPass~\cite{Mato2024}
            & 4.181     & 13.41     & 39.36     & 1.254     & 0.168     & 0.143     & 0.769     & 0.711     & 0.396     \\
            \hline
            TAQR
            & 0.431     & 0.735     & 1.048     & 0.314     & 0.113     & 0.104     & 0.254     & 0.279     & 0.268     \\
            TAQR Adaptive
            & 1.38      & 2.33      & 3.581     & 1.128     & 0.398     & 0.353     & 1.118     & 1.009     & 1.249     \\
            \hline
        \end{tabular}
        \caption{Star transition graph.}
        \label{table:comparison-times-star}
    \end{subtable}

    \vspace{.1cm}

    \begin{subtable}{\textwidth}
        \centering
        \begin{tabular}{l|ccc|ccc|ccc|}
            \multirow{2}{*}{\textbf{Method}}
            & \multicolumn{3}{c|}{$\Gate{X}{}{d}{+1}$} & \multicolumn{3}{c|}{$\Gate{X}{}{d}{-1}$} & \multicolumn{3}{c|}{$\Gate{H}{}{d}{}$} \\
            & \boldmath{$d{=}4$} & \boldmath{$d{=}5$} & \boldmath{$d{=}6$} & \boldmath{$d{=}4$} & \boldmath{$d{=}5$} & \boldmath{$d{=}6$} & \boldmath{$d{=}4$} & \boldmath{$d{=}5$} & \boldmath{$d{=}6$} \\
            \hline
            QSearchPass~\cite{Davis2020}
            & 2446      & 2835      & 4177      & 2524      & 2949      & 4469      & 3159      & 7087      & 11866     \\
            LocQRPass~\cite{Mato2024}
            & 0.352     & 0.489     & 0.651     & 0.371     & 0.553     & 0.775     & 0.761     & 1.27      & 1.823     \\
            LocAdaPass~\cite{Mato2024}
            & 0.872     & 1.65      & 2.437     & 0.953     & 1.565     & 2.339     & 3.658     & 13.0      & 37.24     \\
            \hline
            TAQR
            & 0.286     & 0.439     & 0.576     & 0.262     & 0.387     & 0.47      & 0.441     & 0.736     & 0.966     \\
            TAQR Adaptive
            & 1.203     & 1.952     & 3.11      & 1.213     & 1.821     & 2.542     & 1.57      & 2.568     & 3.822     \\
            \hline
        \end{tabular}
        \begin{tabular}{l|ccc|cccccc|}
            \multirow{2}{*}{\textbf{Method}}
            & \multicolumn{3}{c|}{$\Utry[d]$} & \multicolumn{6}{c|}{\textbf{Two-Qubit Gate}} \\
            & \boldmath{$d{=}4$} & \boldmath{$d{=}5$} & \boldmath{$d{=}6$} & $\Gate{RXX}{}{}{}$ & $\Gate{RZZ}{}{}{}$ & $\Gate{CZ}{}{}{}$ & $\Gate{CX}{}{}{}$ & $\Gate{CH}{}{}{}$ & $\Gate{SWAP}{}{}{}$ \\
            \hline
            QSearchPass~\cite{Davis2020}
            & 3525      & 7992      & 19952     & 2276      & 2289      & 2267      & 2239      & 2229      & 2268      \\
            LocQRPass~\cite{Mato2024}
            & 0.658     & 1.152     & 1.933     & 0.619     & 0.099     & 0.086     & 0.354     & 0.369     & 0.15      \\
            LocAdaPass~\cite{Mato2024}
            & 826.7     & 12.02     & 49.2      & 1.102     & 0.171     & 0.138     & 0.625     & 0.606     & 0.254     \\
            \hline
            TAQR
            & 0.438     & 0.677     & 1.032     & 0.21      & 0.116     & 0.108     & 0.161     & 0.169     & 0.163     \\
            TAQR Adaptive
            & 1.574     & 2.594     & 3.909     & 1.027     & 0.418     & 0.405     & 0.93      & 0.83      & 0.987     \\
            \hline
        \end{tabular}
        \caption{Bipartite transition graph.}
        \label{table:comparison-times-bipartite}
    \end{subtable}

    \caption{Comparison of decompositions produced by given methods in terms of execution time (in milliseconds).
    \TO{} means that the execution exceeded the 1 min timeout.}
    \label{table:comparison-times}
\end{table}

\newpage

\section{Python Code to Execute Decomposition}
\label{app:b}

We use the following Python code to perform a comparison of different methods for qudit unitary decomposition.
Each method produces a qudit circuit format that contains only allowed transition gates and phase-shift gates.

\lstset{
    language=Python,
    backgroundcolor=\color{black!5},
    commentstyle=\color{green!60},
    keywordstyle=\color{blue!60},
    basicstyle=\ttfamily\tiny,
    numbers=left,
    numberstyle=\ttfamily\tiny,
    numbersep=3pt,
    breakatwhitespace=false,
    breaklines=true,
    captionpos=b,
    keepspaces=true,
    showspaces=false,
    showstringspaces=false,
    showtabs=false,
    tabsize=2,
    columns=fixed,
    frame=trbl,
    framesep=1pt,
}

\newsavebox{\bigimage}

\begin{figure}[H]
    \centering

    \sbox{\bigimage}{%
        \begin{subfigure}[b]{.485\linewidth}
            \centering
\begin{lstlisting}[
                caption={Invoke \textit{QSearchPass} decomposition using \texttt{bqskit} package with the version \texttt{1.2.1}.},
                label={listing:code-qsearch},
            ]
import numpy as np
import networkx as nx

from bqskit.ir import Circuit
from bqskit.ir.gates import EmbeddedGate, U1Gate, U1qGate
from bqskit.ir.opt.cost import HilbertSchmidtCostGenerator
from bqskit.ir.opt.minimizers import LBFGSMinimizer
from bqskit.ir.opt.instantiaters import Minimization
from bqskit.qis import UnitaryMatrix
from bqskit.compiler import MachineModel, Compiler, Workflow
from bqskit.passes import SetTargetPass, SetModelPass
from bqskit.passes import QSearchSynthesisPass, SingleQuditLayerGenerator
from bqskit.passes import ScanningGateRemovalPass


class QSearch:
    def decompose(self, utry: np.ndarray, trans: nx.Graph) -> Circuit:
        assert len(utry.shape) == 2 and utry.shape[0] == utry.shape[1]
        d = utry.shape[0]

        target = UnitaryMatrix(utry, [d])

        model = MachineModel(
            1,
            None,
            [EmbeddedGate(U1Gate(), [d], [0, k]) for k in range(1, d)]
            + [EmbeddedGate(U1qGate(), [d], [i, j]) for i, j in trans.edges],
            [d],
        )

        workflow = Workflow(
            [
                SetTargetPass(target),
                SetModelPass(model),
                QSearchSynthesisPass(
                    max_layer=d**2,
                    layer_generator=SingleQuditLayerGenerator(
                        None, allow_repeats=False
                    ),
                    instantiate_options={
                        "method": Minimization(
                            cost_fn_gen=HilbertSchmidtCostGenerator(),
                            minimizer=LBFGSMinimizer(),
                        )
                    },
                ),
                ScanningGateRemovalPass(
                    instantiate_options={
                        "multistarts": 2,
                        "max_iters": 1000,
                        "min_iters": 10,
                    }
                ),
            ]
        )

        with Compiler(num_workers=8, runtime_log_level=40) as compiler:
            circuit = compiler.compile(Circuit(1, [d]), workflow=workflow)
        return circuit
\end{lstlisting}
        \end{subfigure}%
    }

    \usebox{\bigimage}\hfill
    \begin{minipage}[b][\ht\bigimage][s]{.485\linewidth}
        \begin{subfigure}{\linewidth}
            \centering
\begin{lstlisting}[
                caption={Invoke \textit{QSweepPass} decomposition using \texttt{qsweep} package from \url{https://github.com/edyounis/QSweep.git} repository at the commit \texttt{b621547}.},
                label={listing:code-qsweep},
            ]
import numpy as np

from bqskit.ir import Circuit
from bqskit.ir.gates import U1Gate, U1qGate
from bqskit.qis import UnitaryMatrix
from qsweep import QSweepPass


class QSweep:
    def decompose(self, utry: np.ndarray) -> Circuit:
        assert len(utry.shape) == 2 and utry.shape[0] == utry.shape[1]
        d = utry.shape[0]

        target = UnitaryMatrix(utry, [d])
        qsweep_pass = QSweepPass({U1Gate(), U1qGate()})
        return qsweep_pass.synthesize_non_async(target)
\end{lstlisting}
        \end{subfigure}
        \begin{subfigure}{\linewidth}
            \centering
\begin{lstlisting}[
                caption={Invoke \textit{MQT.Qudits} decomposition using \texttt{mqt.qudits} package with the version \texttt{0.4.0}.},
                label={listing:code-mqt},
            ]
import numpy as np
import networkx as nx

from mqt.qudits.core import LevelGraph
from mqt.qudits.compiler import QuditCompiler
from mqt.qudits.quantum_circuit import QuantumCircuit
from mqt.qudits.simulation import MQTQuditProvider
from mqt.qudits.simulation.backends.backendv2 import Backend


class FakeBackend(Backend):
    def __init__(self, trans: nx.Graph, dim: int):
        super().__init__(provider=MQTQuditProvider(), name="FakeBackend")

        graph = LevelGraph(
            edges=trans.edges,
            nodes=list(range(dim)),
            nodes_physical_mapping=list(range(dim)),
            initialization_nodes=[0],
        )
        self._energy_level_graphs: list[LevelGraph] = [graph]

    @property
    def energy_level_graphs(self) -> list[LevelGraph]:
        return self._energy_level_graphs

    def run(self, circuit, **options):
        pass


class MqtQudits:
    def __init__(self, method: str):
        assert method == "LocQRPass" or method == "LocAdaPass"
        self.method = method

    def decompose(self, utry: np.ndarray, trans: nx.Graph) -> QuantumCircuit:
        assert len(utry.shape) == 2 and utry.shape[0] == utry.shape[1]
        d = utry.shape[0]

        circuit = QuantumCircuit(1, [d])
        circuit.cu_one(0, utry)

        backend = FakeBackend(trans, d)
        return QuditCompiler().compile(backend, circuit, [self.method])
\end{lstlisting}
        \end{subfigure}
    \end{minipage}
\end{figure}

\end{document}